\documentclass[twocolumn,aps,prl,showpacs,superscriptaddress,floatfix]{revtex4-1}
\usepackage[dvips]{color} 
\usepackage{graphicx} 
\usepackage{bm}

\begin{document}

\title{Fission decay of $^{282}$Cn studied using cranking inertia}

\author{D. N. Poenaru}
\email[]{poenaru@fias.uni-frankfurt.de}
\author{R. A. Gherghescu}
\affiliation{Horia Hulubei National Institute of Physics and Nuclear
Engineering (IFIN-HH), \\P.O. Box MG-6, RO-077125 Bucharest-Magurele, Romania}
\affiliation{Frankfurt Institute for Advanced Studies (FIAS),
Ruth-Moufang-Str. 1, 60438 Frankfurt am Main, Germany}

\date{ }

\begin{abstract}
Superheavy nuclei produced until now are decaying mainly by
$\alpha$~emission and spontaneous fission.  Calculated $\alpha$~decay
half-lives are in agreement with experimental data within one order of
magnitude.  The discrepancy between theory and experiment can be as high as
ten orders of magnitude for spontaneous fission.  We analyze a way to
improve the accuracy by using the action integral based on cranking inertia
and a potential barrier computed by the macroscopic-microscopic method with
a two-center shell model. Illustrations are given for $^{282}$Cn which has a
measured fission half-life.
\end{abstract}


\maketitle

\section{Introduction}

The heaviest superheavy (SH) elements \cite{ham13arnps,khu14prl,sob11ra}
with atomic numbers $Z=107 - 112$ are produced in cold fusion reactions
\cite{hof00rmp,mor07jpsjb} between targets $^{208}$Pb or $^{209}$Bi and
beams with mass numbers $A > 50$; hot fusion reactions between actinide
nuclei and $^{48}$Ca \cite{oga07jpg} are used for those with $Z=113 - 118$. 
Attempts to study $Z=120$ are reported \cite{oga09pr,hof11pc,khu14prl}.  The
main decay modes are either $\alpha$~decay or spontaneous fission. 
Spontaneous fission, the dominating decay mode in the region around Rf,
becomes a relatively weaker branch compared to $\alpha$-decay for the
majority of recently discovered proton-rich nuclides.  According to our
calculations \cite{p309prl11,p315prc12,p311jpg12} it would be possible to
see also cluster~decay (CD) \cite{enc95,ps84sjpn80} for heavier SHs with $Z
> 121$, unlike for $Z=87-96$, where CD is a rare phenomenon in a huge
background of alpha particles.  Up to now the produced SHs are
neutrondeficient nuclei far from the line of $\beta$-stability.  In the
future it is expected to synthesize SHs closer to $\beta$-stability line
\cite{zag13prc}.

For spontaneous fission calculations we refer to Hartree-Fock-Bogoliubov
approach with finite-range and density-dependent Gogny force \cite{war12prc}
and self-consistent symmetry-unrestricted nuclear density functional with
Skyrme energy density functional and cranking inertia \cite{sta13prc}.  The
closest value to the experimental one was obtained by using a dynamical
approach \cite{smo97pr,smo95pr}.  Simple relationships
\cite{bao13np,san10npa,xu08prc} have also been used.

Previously we have shown \cite{p326jpg13} that calculated $\alpha$~decay
half-lives are in agreement with experimental data within one order of
magnitude, while the discrepancy between theory and experiment can be as
high as ten orders of magnitude for spontaneous fission.  It was clear that
Werner-Wheeler approximation \cite{pg197pr95,pg264pr05} for the nuclear
inertia leads to too small values to explain the measured spontaneous
fission half-life.  In the present work we try a better method based on the
microscopic cranking inertia \cite{pg331ep14,sch86zp,p261epja05} introduced
by Inglis \cite{ing54pr}.  Deformation energy and the inertia tensor are
calculated in order to determine the half-life.  Potential barrier is
computed by the macroscopic-microscopic method \cite{str67np} with a
two-center shell model \cite{ghe03prc}.

\section{Deformation Energy}

In a spontaneous fission process we obtain a light, $^{A_2}Z_2$, and a
heavy, $^{A_1}Z_1$, fragment with radii $R_2$ and $R_1$ from a parent nucleus
, $^{A}Z$, with a radius $R_0$:
\begin{equation}
^{A}Z \ \rightarrow \ ^{A_1}Z_1 + ^{A_2}Z_2
\end{equation}

According to the macroscopic-microscopic method  the deformation energy of a
nucleus, $E_{def}$, is calculated as a sum of two terms coming from a
phenomenological (e.g.  Yukawa-plus exponential model (Y+EM)),
$E_{macro}=E_{Y+EM}$, and a small shell-plus-pairing correction, $E_{micro} =
\delta E$: 
\begin{equation}
E_{def} = E_{Y+EM} + \delta E
\end{equation}
The shape-dependent terms in the LDM are the surface energy due to the
strong interactions, tending to hold the nucleons together, and the   
electrostatic (Coulomb) energy, acting in the opposite direction. 
By requesting zero deformation energy for a spherical shape,
\begin{equation}
E_{Y+EM} = (E_Y - E_Y^0)+(E_c - E_c^0) 
\end{equation}
the Coulomb energy and the nuclear energy are both expressed as
double-volume integrals:
\begin{equation}
E_Y =- \frac {a_2}{8\pi ^2 r_0^2 a^4} \int_{V_n} \int \left ( \frac
{r_{12}}{a}-2 \right ) \frac {\exp (- r_{12}/a)}{ r_{12}/a} 
d^3 r_1 d^3 r_2
\label{yuka}
\end{equation}
where $r_{12}=|{\bf r}_1 - {\bf r}_2|$, $a$ is the diffusivity parameter,
and $a_2 = a_s(1-\kappa I^2)$, $I=(N-Z)/A$.
\begin{equation}
E_c = \frac {1}{2} \int_{V_n} \int \frac {\rho _e ({\bf r}) 
\rho _e ({\bf r}_1) d^3 r d^3 r_1}{|{\bf r} - {\bf r}_1|}
\label{ecoul}
\end{equation}
with $\rho _e $ the charge density.
\begin{figure}[htb] \includegraphics[width=6.6cm]{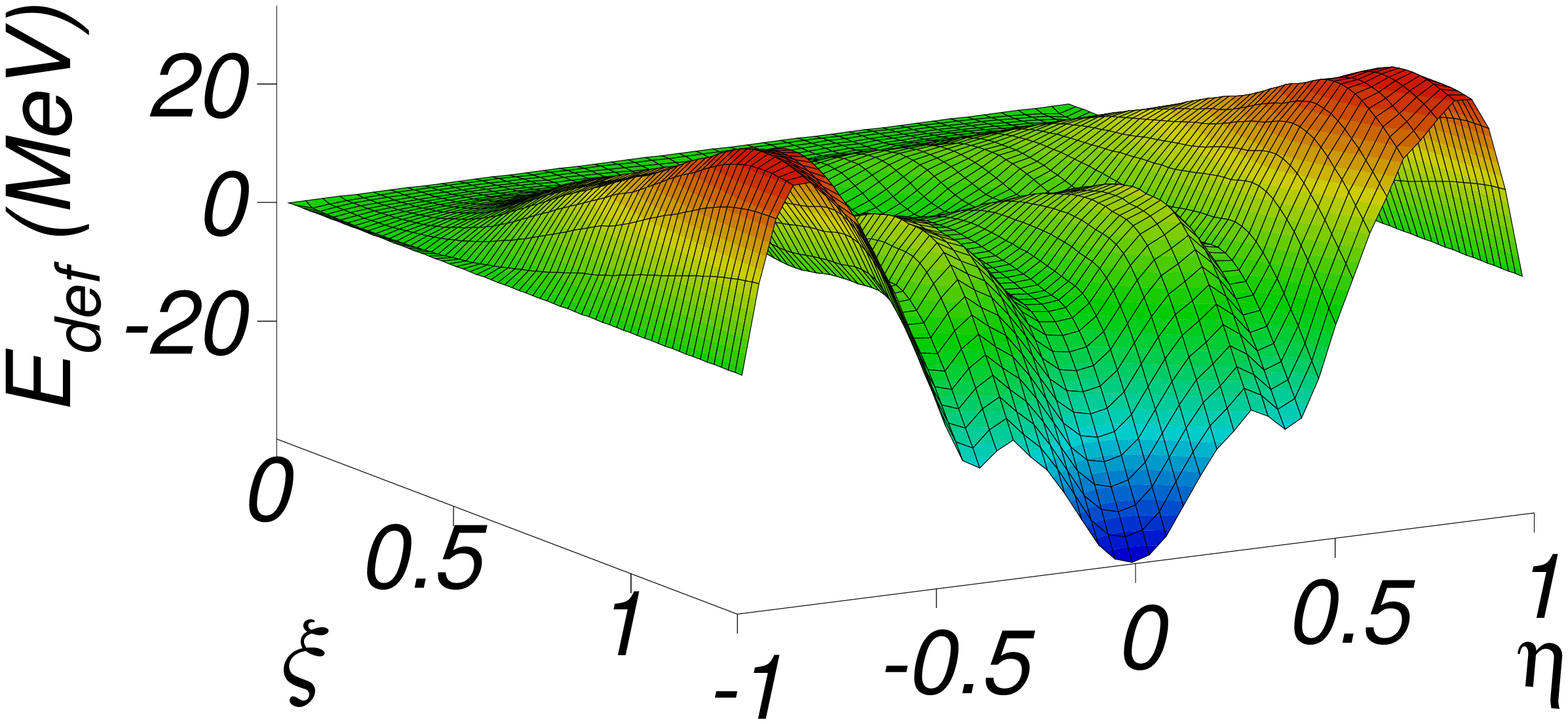} 
                    \includegraphics[width=6.6cm]{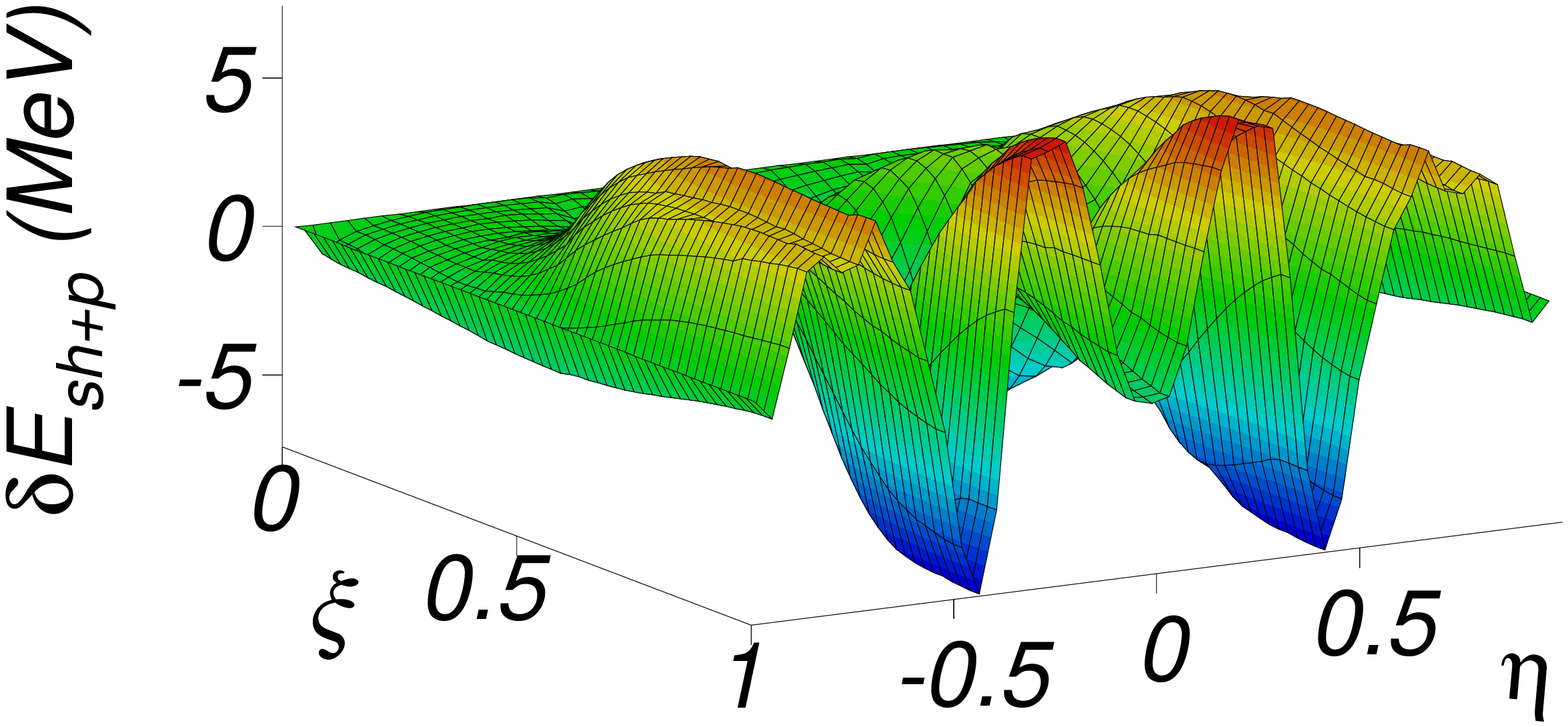} 
                    \includegraphics[width=6.6cm]{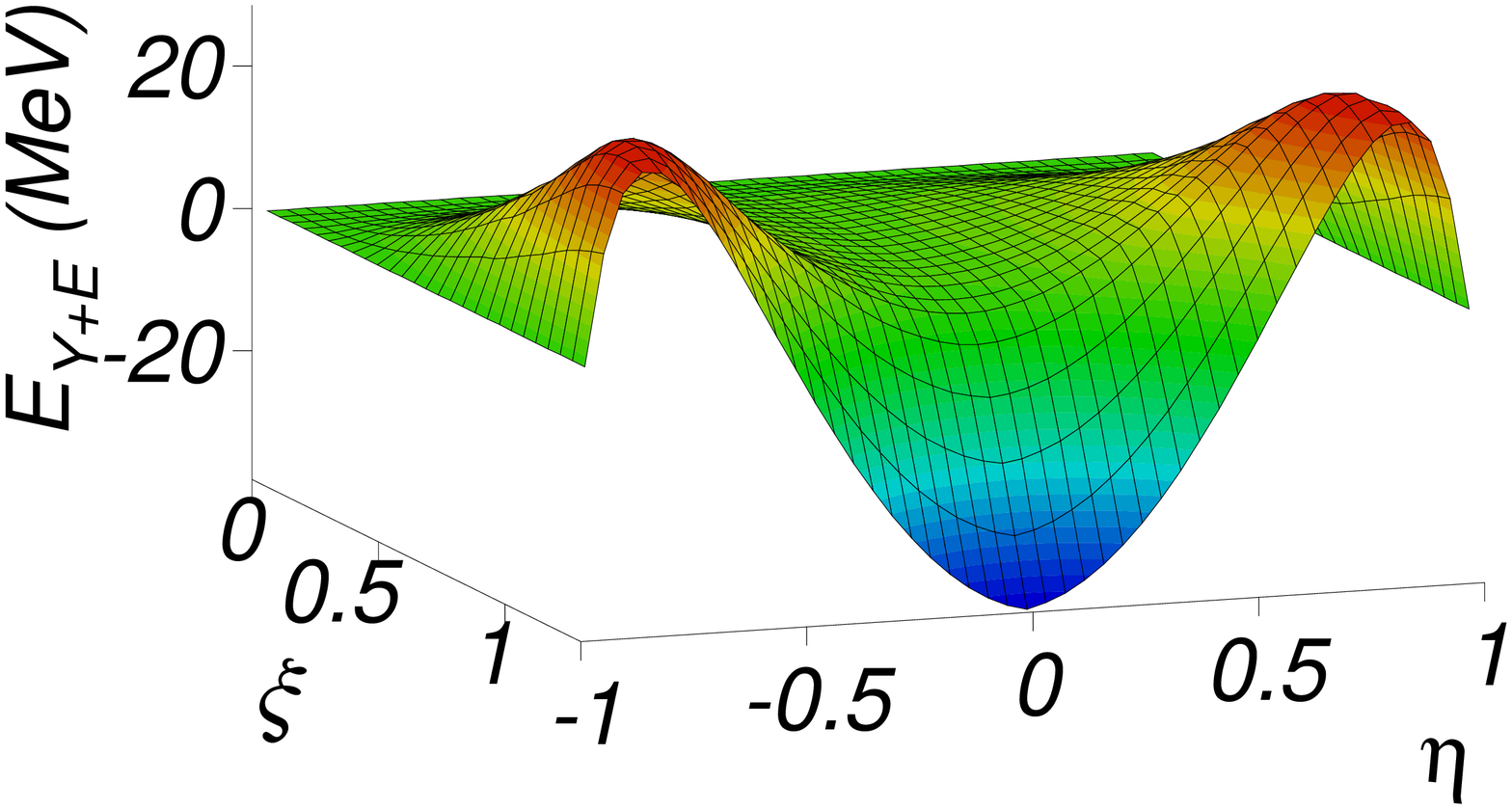} 
\caption{(Color online) PES of $^{282}$Cn vs $\xi$ and $\eta$.  Y+EM
(bottom), shell + pairing corrections (center), and total deformation energy
(top).  The deepest two valleys in the central figure are due to the
magicity of the daughter $^{208}$Pb for $^{74}$Zn radioactivity at about
$\eta =0.47$ and of the light fragment $^{132}$Sn at $\eta \simeq 0$. 
\label{pescn}} 
\end{figure}

\begin{figure}[htb] 
\includegraphics[width=7.6cm]{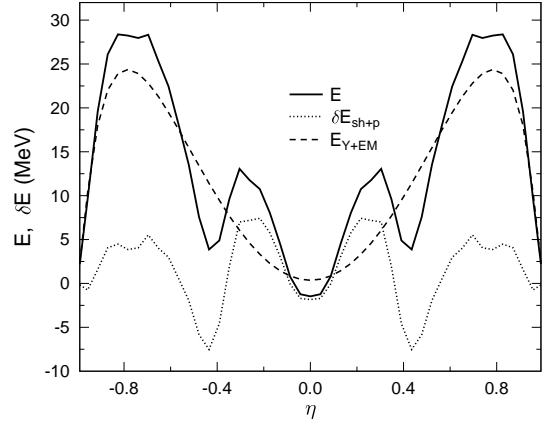} 
\caption{(Color online) Deformation energies at the touching point
configurations ($R=R_t$) of $^{282}$Cn vs the asymmetry $\eta $: macroscopic
energy $E_{Y+EM}$; shell + pairing corrections $\delta E_{sh+p}$ and their
sum $E=E_{def}$.
\label{tou}} 
\end{figure}

Potential energy surfaces \cite{p266pr06} are calculated by using the most
advanced asymmetric two-center shell model allowing to obtain shell and
pairing corrections which are added to the Yukawa-plus-exponential model
deformation energy \cite{kra79pr} taking into account the difference between
charge and mass asymmetry \cite{p80cpc80}.  The model parameters are taken
from M\" oller et al.~\cite{mol95adnd}.

Starting from the touching point configuration, $R \geq R_t=R_1 + R_2$, for
spherical shapes of the fragments, one can use {\em analytical
relationships.} The Coulomb interaction energy of a system of two spherical
nuclei, separated by a distance $R$ between centers, is $E_{c12} = e^2 Z_1
Z_2 /R$, where $e$ is the electron charge.

Within a liquid drop model (LDM) there is no contribution of the surface
energy to the interaction of the separated fragments; the barrier has a
maximum at the touching point configuration.  The proximity forces acting at
small separation distances (within the range of strong interactions) give
rise in the Y+EM to an interaction term expressed as follows
\begin{eqnarray}
E_{Y12} =& -4\left ( \frac{a}{r_0} \right ) ^2 \sqrt {a_{21} a_{22}}
\frac{\exp (- R/a)}{R/a} \nonumber \\  & \cdot
\left [ g_1 g_2 \left ( 4+\frac{R}{a} \right ) -g_2f_1 - g_1f_2
\right ]
\end{eqnarray}
where
\begin{equation}
g_k =  \frac{R_k}{a} \cosh \left ( \frac{R_k}{a}  \right ) - \sinh
\left ( \frac{R_k}{a} \right ) 
\end{equation}
\begin{equation}
f_k = \left (\frac{R_k}{a}
\right ) ^2 \sinh \left ( \frac{R_k}{a} \right )
\end{equation}

In order to obtain a relatively smooth potential energy surface (PES) we
made the approximation for mass and charge asymmetry
$\eta=\eta_A=(A_1-A_2)/A \simeq \eta_Z=(Z_1-Z_2)/Z$.  We prefer to use the
dimensionless separation distance $\xi=(R-R_i)/(R_t-R_i)$ instead of $R$. 
Here $R_i=R_0-R_2$.  In this way one can clearly see the touching point
configuration at $\xi=1$.  We also adopt the usual convention of having zero
deformation energy and shell plus pairing corrections for the initial
spherical shape, leading to $E_{def} = E_{Y+EM} = \delta E =0$ at $R=R_i$
for all values of $\eta$ and at $\eta =\pm 1$ for all values of $R$. 

The PES of $^{282}$Cn versus the normalized separation distance $\xi$ and
the mass asymmetry $\eta$ are plotted in Fig.~\ref{pescn}.  The macroscopic
Y+EM deformation energy is shown at the bottom, followed by the microscopic
shell plus pairing corrections (center), and their sum (the total
deformation energy) at the top.  Two valleys around $|\eta | \simeq 0$ and
$0.47$ can be seen in the center of Fig.  \ref{pescn} as well as in
Fig.~\ref{tou} at the touching point $R=R_t$.  They are produced due to the
magicity of the nucleon number of one of the fragments around $^{132}$Sn and
$^{208}$Pb, respectively.  
\begin{figure*}[htb] \includegraphics[width=10.6cm]{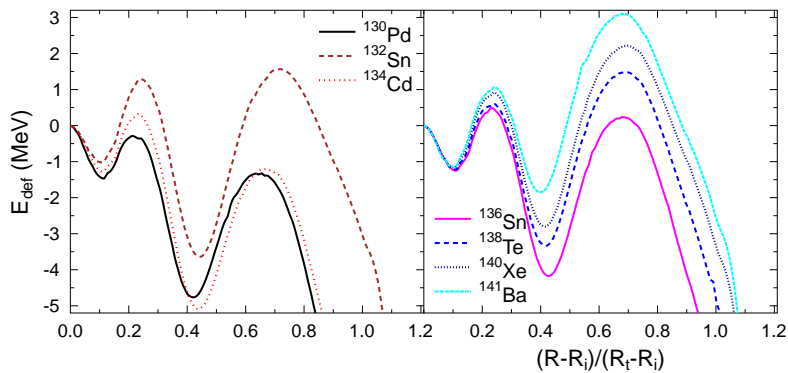} 
\caption{(Color online) Potential barriers for the spontaneous fission of
$^{284}$Cn with light fragments $^{130}$Pd, $^{132}$Sn, $^{134}$Cd (left)
and $^{136}$Sn, $^{138}$Te, $^{140}$Xe, $^{141}$Ba (right).  \label{7bar}}
\end{figure*} 
Every time a nucleon number reaches a magic value, the corresponding shell
correction has a local minimum.

Few fission channels could be efficiently used to test the method of
calculating spontaneous fission half-lives of $^{282}$Cn who's experimental
value is known.  We know from the mass distributions of fission fragments
that few splittings give the largest yield, very likely explained by shell
effects (spherical and/or deformed magic numbers of neutrons and protons). 
Consequently we can guess that the channels having the light fragments shown
in Fig.~\ref{7bar} would be among those major splittings.

\section{Nuclear Inertia and the Half-life}

After including the BCS pairing correlations \cite{bar57pr}, the 
components of the inertia tensor is given by
\cite{bra72rmp}:
\small{\begin{equation}
B_{ij} =2\hbar^2\sum_{\nu \mu} \frac{\langle \nu|\partial H/\partial
\beta_i|\mu \rangle \langle \mu|\partial H/\partial \beta_j|\nu
\rangle}{(E_\nu +E_\mu)^3}(u_\nu v_\mu +u_\mu v_\nu)^2 
\normalsize
\label{eq3}
\end{equation}
where $H$ is the single-particle Hamiltonian allowing to determine the
energy levels and the wave functions $|\nu \rangle$; $u_\nu$, $v_\nu$ are
the BCS occupation probabilities, $E_\nu$ is the quasiparticle energy. 
Other involved quantities are the pairing gap $\Delta$ and the Fermi energy
$\lambda$ \cite{p318rrp12}.  The multidimensional hyperspace of deformation
parameters is defined by $\beta_1, \beta_2, ...., \beta_n$.  The dimension
of any component, $B_{ij}$, of the tensor is a mass.  By choosing the
distance between fragments, $R$, as deformation coordinate, the effective
mass at the touching point of the two fragments should be equal to the
reduced mass $\mu = (A_1A_2/A)m$, where $m$ is the nucleon mass.  Sometime
the inertia tensor is called a mass tensor.  Similar to the shell correction
energy, the total inertia is the sum of contributions given by protons and
neutrons $B_{ij}=B_{ij}^p+B_{ij}^n$.

\begin{figure}[htb] \includegraphics[width=8.6cm]{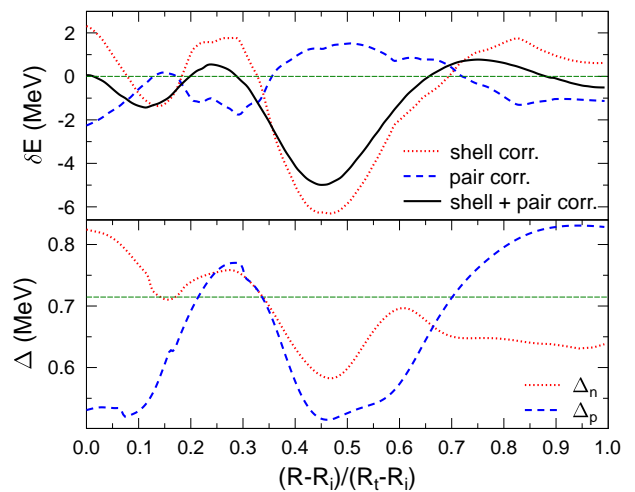} 
\caption{(Color online) Shell and pairing corrections (top), pairing gap
(bottom) versus $(R-R_i)/(R_t-R_i)$ for the fission of $^{282}$Cn into
$^{130}$Pd + $^{152}$Dy fragments.  Proton and neutron component are shown.
\label{pair}} \end{figure}
The kinetic energy 
\begin{equation} 
E_k = \frac{1}{2}\sum_{i,j=1}^n B_{ij}(\beta)\frac{d\beta_i}{dt} 
\frac{d\beta_j}{dt} 
\end{equation} 
includes the change in time of the nuclear shape (the time derivatives).

By choosing four independent deformation parameters $R, b_2, \chi _1,\chi
_2$ \cite{pg264pr05} during the deformation from one parent nucleus to two
fission fragments, the surface equation in cylindrical coordinates $\rho ,
z$ is given by: 
\begin{equation}
\rho _s^2(z;b_1,\chi _1,b_2,\chi _2)= \left \{ \begin{array}{ccc}
b_1^2-\chi _1 ^2z^2 & , -a_1 < z < z_c \\
b_2 ^2 -\chi _2 ^2 (z-R)^2 & , z_c<z<R+a_2
\end{array} \right.
\end{equation}
where $z_c$ is the position of the crossing plane.
\begin{figure}[htb]
\includegraphics[width=8.6cm]{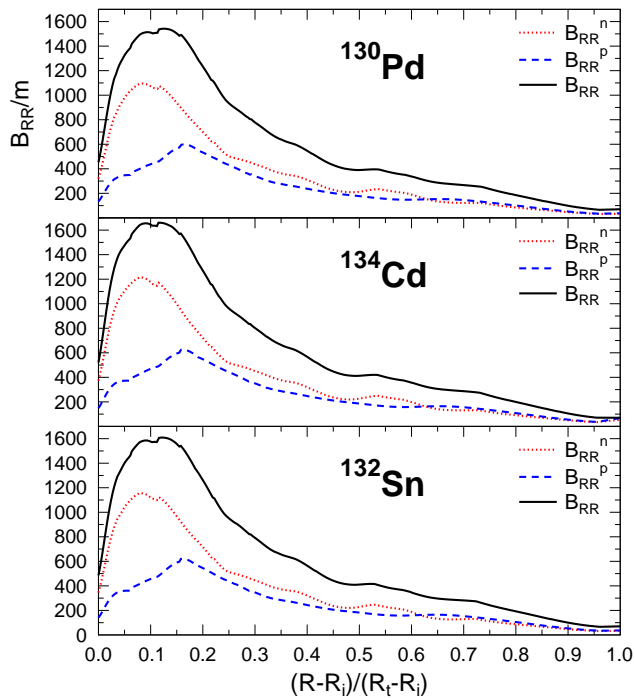} 
\caption{(Color online) The component B$_{RR} /m$ of the inertia tensor 
versus $(R-R_i)/(R_t-R_i)$ for the fission of $^{282}$Cn with light
fragments $^{130}$Pd, $^{134}$Cd, and $^{132}$Sn.  
Proton and neutron components are shown. 
\label{brr4}}
\end{figure} 
The semiaxes ratio of spheroidally deformed fragments are denoted by $\chi
_1= b_1/a_1$, $\chi _2=b_2/a_2$.  The scalar, $B(R)$, is determined by the
components of the tensor and the partial derivatives with respect to $R$:
\begin{eqnarray} 
B(R) & = & B_{b_2b_2}\left( \frac{db_2}{dR} \right)^2+
2B_{b_2 \chi _1}\frac{db_2}{dR}\frac{d\chi _1}{dR} +2B_{b_2 \chi _2}
\frac{db _2}{dR} \frac{d\chi _2}{dR}+ \nonumber \\ & & 2B_{b _2 R}\frac{db
_2}{dR} + B_{\chi _1 \chi _1}\left(\frac{d\chi _1}{dR}\right)^2+ 2B_{\chi _1
\chi _2}\frac{d\chi _1}{dR}\frac{d\chi _2}{dR}+ \nonumber \\ & & 2B_{\chi _1
R}\frac{d\chi _1}{dR}+B_{\chi _2 \chi _2} \left(\frac{d\chi
_2}{dR}\right)^2+ 2B_{\chi _2 R}\frac{d\chi _2}{dR}+ B_{RR} 
\end{eqnarray}
One of the most important binary split of $^{282}$Cn leads to $^{130}$Pd +
$^{152}$Dy fragments.  As it is shown in Fig.~\ref{pair}, the variation of
the shell plus pairing correction (top) determines the pairing gaps,
$\Delta_n$ and $\Delta_p$ (bottom), as solutions of the BCS system of two
eqs.  also allowing to find the Fermi energies $\lambda_n$ and $\lambda_p$
\cite{p195b96c,p318rrp12}.  Their influence on the inertia components
B$_{RR} ^n/m$ and B$_{RR}^p /m$ is also clear from the Figure~\ref{brr4}. 
B$_{RR}$ is the most important component of the inertia tensor if we
consider spherical shapes and keep constant the radius $R_2$ of the light
fragment.  When we compare B$_{RR}/m$ for the binary fission $^{282}Cn \
\rightarrow \ \ ^{130}Pd + ^{152}Dy$ (see the top of Fig.~\ref{brr4}) with
those corresponding to $^{282}Cn \ \rightarrow \ \ ^{134}Cd + ^{148}Gd$
(center of Fig.~\ref{brr4}) and to that of $^{282}$Cn fission with light
fragment $^{130}$Sn (bottom of Fig.~\ref{brr4}) it is clear that the first
one is smaller than the two others.  Having also in mind a broader potential
barrier we may expect a longer half-life for this split.  The three curves
are very similar as it should be due to the fact that the shell plus pairing
effects are not very much different for these neigbouring light fragments
($Z_2, N_2$ (46, 84) for $^{130}$Pd, (50,82) for $^{132}$Sn, and (48, 86)
for $^{134}$Cd).  In all three cases neutron contribution is larger than the
proton one.  Intuitively we can guess that an exponential approximation
would be appropriate in the interval of the variable $\xi$ between the two
turning points.

The half-life of a parent nucleus $AZ$ against the split into a light
fragment $A_2Z_2$ and a heavy fragment $A_1 Z_1$ is given by
\begin{equation}
T = [(h \ln 2)/(2E_{v})] exp(K_{ov} + K_{s})
\end{equation}
and is calculated by using the Wentzel--Kramers--Brillouin (WKB) 
quasiclassical approximation, according to
which the action integral is expressed as
\begin{equation}
K=\frac{2\sqrt{2m}}{\hbar}\int_{R_a}^{R_b}
\{[(B(R)/m)][E_{def}(R)-E_{def}(R_a)]\}^{1/2}dR
\end{equation}
with $B=$ the cranking inertia, $K=K_{ov}+K_s$, and the $E(R)$ potential 
energy of deformation. $R_a$ and $R_b$ are the turning points of the WKB
integral where $E_{def} = E_{def}(R_a) = E_{def}(R_b) $.
The two terms of the action integral $K$, correspond to the overlapping
($K_{ov}$) and separated ($K_s$) fragments. We can use the relationship
\begin{equation}
\log_{10} T = 0.43429(0.4392158S_{ab}) -20.8436 - \log_{10} E_v
\end{equation}
where
\begin{equation}
S_{ab} = \int _{R_a}^{R_b} \{[(B(R)/m)][E_{def}(R)-E_{def}(R_a)]\}^{1/2} dR
\end{equation}

\begin{table}[hbt] 
\caption{Decimal logarithm of fission half-lives $\log_{10} T_f
(s)$ of $^{282}$Cn with different light fragments and zero point vibration
energies. Experimental value: $\log_{10} T_f^{exp}(s)=-3.086$ 
\label{tab}} 
\begin{center}
\begin{ruledtabular}
\begin{tabular}{cccccc}
Light fragment & $E_v$ (MeV)&  $\log_{10} T_f (s)$ \\
\hline
$^{130}$Pd     & 0.50000 & -3.4278 \\
               & 0.43680 & -3.0860 \\

$^{134}$Cd     & 0.50000 & -2.4881 \\
               & 0.60289 & -3.0860 \\

$^{132}$Sn     & 0.50000 &  5.5076  \\
               & 1.96620 & -3.0860 \\

\end{tabular}
\end{ruledtabular}
\end{center}
\end{table}

The computer programme developed by one of us (RAG) to calculate the
half-life starts by making a search for the largest barrier height near the
deepest minimum (the ground-state). The two turning points are found in the
next step. Usually for superheavy nuclei the second turning point
corresponds to separated fragments inside, $R_b < R_t$, or outside, $R_b >
R_t$, the touching point.

In Table~\ref{tab} we present the results of our calculations concerning the
half-life for spontaneous fission of $^{282}$Cn.  It is clear that an
important binary split will produce the light fragment $^{130}$Pd --- the
same giving the thinnest (most penetrable) potential barrier in
Fig.~\ref{7bar}.  On the other hand when the light fragment is $^{134}$Cd or
$^{132}$Sn the half-life is longer for the same zero-point vibration energy. 
We can conclude that by using the cranking inertia it is possible to
reproduce the spontaneous fission half-life with a reasonable value of the
zero-point vibration energy.

\begin{acknowledgments}
This work is partially supported within the IDEI Programme under contracts
43/05.10.2011 and 42/05.10.2011 with UEFISCDI, and NUCLEU Programme,
Bucharest.
\end{acknowledgments}


\end{document}